\documentstyle[epsfig,psfig,aps]{revtex}
\begin{document}
\title{Nuclei, Superheavy Nuclei and Hypermatter in a chiral $SU(3)$-Model}
\author{\underline{Ch. Beckmann}
\footnote{Christian Beckmann, Institut f\"ur Theoretische
Physik, Universit\"at Frankfurt am Main, Robert-Mayer-Str. 8-10, 
60325 Frankfurt am Main, Germany, Tel.:+49-(0)69-7982-2630, Fax:-8350,
e-Mail: beckmann@th.physik.uni-frankfurt.de},
P. Papazoglou, D. Zschiesche\\ 
S. Schramm, H. St\"ocker, W. Greiner\\ {\it Institut f\"ur 
Theoretische Physik\\
Johann Wolfgang Goethe Universit\"at, D-60054 Frankfurt
am Main}\\
}
\maketitle
\newcommand{\be}{\begin{equation}}
\newcommand{\ee}{\end{equation}}
\newcommand{\bea}{\begin{eqnarray}}
\newcommand{\eea}{\end{eqnarray}}
\newcommand{\pa}{\partial}
\newcommand{\pasl}{\partial \!\!\! /}
\newcommand{\eh}{\frac{1}{2}}
\newcommand{\li}{\left}
\newcommand{\re}{\right}
\newcommand{\ovl}{\overline}
\newcommand{\dint}{{\mathrm d}}
\newcommand{\no}{\nonumber}
\newcommand{\q}{\quad}
\newcommand{\e}{{\mathrm e}}
\newcommand{\im}{{\mathrm i}}

\begin{abstract}
A model based on chiral $SU(3)$-symmetry in nonlinear realisation is
used for the investigation of nuclei, superheavy nuclei, hypernuclei
and multistrange nuclear objects (so called MEMOs).
The model works very well in the case of 
nuclei and hypernuclei with one $\Lambda$-particle and rules out MEMOs.
Basic observables which are known for nuclei and hypernuclei
are reproduced satisfactorily.
The model predicts $Z$=120 and $N$=172, 184 and 198 as the next
shell closures in the region of superheavy nuclei.
The calculations have been performed in self-consistent relativistic
mean field approximation assuming spherical symmetry.
The parameters were adapted to known nuclei.
\end{abstract}
\subsection{Introduction}
The theory of strong interactions (QCD) is currently still unsolvable
in the nonpertubative low-energy regime.
Therefore different approaches have to be found for predicting 
nuclear properties.
A promising ansatz are the effective hadronic field theories.
Extensive studies of nuclei and hypernuclei have been done in this field.
The probably most common effective models are relativistic mean 
field models \cite{walecka}.
Important work has been done for example by S.\ \'Cwiok et al. \cite{cwiok} 
as well as K.\ Rutz, J.\ Maruhn, P.-G. Reinhard et al.\ (e.g. \cite{rut97}).
Also hypermatter has been investigated up to extreme objects with 
various combinations of
hyperons (e.g. \cite{mil88,ine96,mar89,ruf88,sch92,sch93,sch94a,sch94b}).\\
In the present work we expand this ansatz. Here we
develop a model closer to the symmetry of the 
elementary theory, i.e. QCD.
Therefore we construct a Lagrange density, which is
largely chiral symmetric, similar to the linear sigma-model \cite{gellmann}.
The chiral symmetry is indeed broken explicitly to give the 
pseudoscalar mesons, the goldstone bosons,
a mass and to model the larger finite mass of the strange quark
in the hyperon sector.\\
The model has already been applied successfully to infinite nuclear 
matter \cite{paper3}.\\
The present paper presents the investigations of finite nuclei, superheavy
nuclei and hypernuclei with one and more hyperons in this framework.
\subsection{The model}
For our investigations we introduce an effective model based on chiral 
symmetry in nonlinear realisation.
The nonlinear realisation of chiral symmetry has been chosen because
in the linear version \cite{paper2} the coupling of the spin-0-mesons to 
the baryons is restricted to the symmetric coupling (d-type), 
while in accordance with the vector-meson-dominance 
the vector mesons couple antisymmetrically to the baryons.
The result is a disturbance of the balance between the repulsion of the 
vector field and the attraction of the scalar field
in the nucleonic potential.
There is a natural coupling of the nucleon to the strange scalar ($\zeta$) 
field but not to the corresponding strange vector field $\phi$.
As a result the scalar attraction is not
compensated by an adequate vector repulsion. \\
We will now discuss our specific Lagrangian.
For our calcuations we worked in the mean-field-approximation, that
means quantum fluctuations of the meson fields are neglected and 
the quantum fields are replaced by their expectation values:
\bea
\omega&=&\langle \omega_0 \rangle \nonumber \\
\rho&=&\langle \rho_0 \rangle 
\eea 
In case of the vector mesons ($\rho$, $\omega$, $\phi$), 
the space like parts vanish due to isotropy
and only the time component survives.
The pion-expectation value vanishes
because of good parity in the groundstate.
Furthermore we adopt the no-sea-approximation and 
perform all calculations in spherical symmetry.\\
Finally we get the following Lagrangian:

\newcommand{\cL}{{\cal L}}
\bea
\label{mflagrdichte}
\cL_{\mathrm kin} &=& i\sum\limits_i\bar{B_i}\gamma_i\partial^\mu\!
B_i-\frac{1}{2}\!\!\sum
\limits_{\Phi=\sigma, \zeta, \chi, \omega, \phi, \rho, A}
\!\!\partial_\mu\Phi\partial^\mu\!\Phi \nonumber\\
&&\nonumber\\
\cL_{\mathrm int}&=&-\sum\limits_i\bar{B_i}\gamma_0
\left[g_{i_\omega}\omega_0
+g_{i\rho}\tau_3\rho_0+g_{i\phi}\phi_0 
+\frac{1}{2}e(1+\tau_3)A_0+m_i^\ast\gamma_0\right]B_i\nonumber\\
&&\nonumber\\
\cL_{\mathrm vec}&=&-\frac{1}{2}k_0 \frac{\chi^2}{\chi_0^2}\left(
m_\omega^2\omega^2+m_\rho^2\rho^2\right)
+  g_4^4\left(\omega^4+6\omega^2\rho^2+\rho^4\right)\nonumber\\
&&\nonumber\\
\cL_0^{\mathrm{chi}}& =& -\frac{1}{2}k_0\chi^2(\sigma^2+\zeta^2)+ k_1(\sigma^2+\zeta^2)^2
+k_2\left(\frac{\sigma^4}{2}+\zeta^4\right)\nonumber\\
&& + k_3\chi\sigma^2\zeta
+k_{3m}\chi\left(\frac{\sigma^3}{\sqrt{2}}+\zeta^3\right)
-k_4\chi^4\nonumber\\
&& -\frac{1}{4}\chi^4\mbox{ln}\frac{\chi^4}{\chi_0^4}
+\frac{\delta}{3}\chi^4\mbox{ln}\frac{\sigma^2\zeta}{\sigma_0^2\zeta_0}\nonumber\\
&&\nonumber\\
\cL_{\mathrm SB}& =& -\left(\frac{\chi}{\chi_0}\right)^2\left[x\sigma+y\zeta\right]
\eea
with
 $x=m_\pi^2 f_\pi$ and $y=\sqrt{2}m_{\mathrm K}^2 f_{\mathrm K} 
-\frac{1}{\sqrt{2}}m_\pi^2 f_\pi$.
The fields $B_i$ represent the baryons considered, i.e.
nucleons in the case of normal nuclei and furthermore 
hyperons for the investigation of hypernuclei.\\
The incorporated mesons are the scalar non-strange ($\sigma$) and strange ($\zeta$) mesons,
the vector mesons $\omega$, $\phi$ and $\rho$
and the coulomb-field $A$. Furthermore a scalar glueball field $\chi$ 
has been taken into account
for the broken
scale invariance \cite{paper3}. 
The second part of the Lagrangian exhibits the interaction between
baryons and mesons and the photon.
Here $m_i^\ast$ is the effective mass of the baryon species $i$  which is
generated by the interaction with the scalar mesons:
\be
m_i^\ast =g_{i\sigma}\sigma+g_{i\zeta}\zeta\quad .
\ee
The asterisk indicates that the masses in the medium shift with the changing 
scalar fields $\sigma$ and $\zeta$.
The third part ${\cal L}_{\mathrm vec}$ contains the vector meson mass terms
as well as a quartic vector meson self interaction.
The term ${\cal L}_0$ indicates the potential of the scalar fields of the model.
The logarithmic terms in the potential break the scale invariance {\cite{paper3}}.\\
The spontanously broken symmetry leads to the existence of massless
Goldstone modes which are eliminated from the theory by
the last term of the Lagrangian, which explicitly breaks the chiral
symmetry.\\
For the finite-nucleus calculation we also considered pairing effects in a 
simple approximation.
The pairing was parametrized in an isospin symmetric way with
the {\it Constant-Gap-Model} \cite{blo76}.

\subsection{Adjustment of the Parameters}
The free parameters of the model have been adjusted to the properties of
finite nuclei in a least square fit.
The function to minimize is defined as
\be
\chi^2=\sum\limits_n\li(\frac{O^{\mathrm exp}_n-O^{\mathrm theo}_n}{\Delta O_n}\re)^2\q ,
\ee
where $O^{\mathrm exp}$ are the experimental 
and $O^{\mathrm theo}$ the calculated values of the observables.
The value $\Delta O$ is used to weight the different observables.\\
The following nuclei enter into the fit:
\be
^{16}{\mathrm O}, ^{40}{\mathrm Ca}, ^{48}{\mathrm Ca}, ^{58}{\mathrm Ni}, 
^{90}{\mathrm Zr},  ^{112}{\mathrm Sn},  ^{124}{\mathrm Sn},  ^{208}{\mathrm Pb}\q.
\ee
These nuclei are spherical and the excited states are not too close to 
the ground state, so their influence is minimal.
The observables that are relevant for the fit are
the binding energy $E_{\mathrm B}$,
the charge diffraction radius $R_{\mathrm diff}$,
the surface thickness $\sigma_{\mathrm O}$ and
in some cases the binding energy of $^{264}{\mathrm Hs}$,
as discussed below.
The values of the observables used for the fit have been taken from
\cite{fri86}, the minimization procedure used is Powell's method \cite{numrec}.
\subsection{The Parameters}
The parameters, varied in the $\chi^2$-fit, are
the coupling constant between the nucleons and the $\omega$-meson $g_{{\mathrm N}\omega}$,
the coupling constant between the nucleons and the $\rho$-meson $g_{{\mathrm N}\rho}$,
which primarily determines the isospin dependent part of the nuclear force,
the vacuum expectation value of the gluon condensate $\chi_0$,
the decay constant of the kaon $f_{\mathrm K}$,
the coupling constant of the quartic self interaction of the $\omega$-vector meson $g_4$,
and the parameters of the potential of the scalar mesons $k_1$ and $k_2$.
The absolute values of the parameters and their change in the fit
are shown in Table \ref{parameter}.
For the set $C_1$ the parameters were adjusted to the vacuum properties of
the mesons as well as to the properties of infinite nuclear matter \cite{paper3}.
The set $C_1^{\mathrm nuc}$ has been derived from a fit to finite nuclei
as outlined above. 
We have also performed a fit suitable for extrapolating
superheavy nuclei, which includes
the binding energy of $^{264}$Hs but disregards the observables of $^{16}$O.
The resulting parameter set is called $C_1^{\mathrm hs}$.\\
To undestand the significance of the variations in the parameters,
we consider the sensitivity of $\chi^2$ with respect to the
parameters.
One finds that $\chi^2$ is very sensitive to
the value of $g_{{\mathrm N}\omega}$:
a change of $g_{{\mathrm N}\omega}$ of 0.25 \% leads to a variation of
$\chi^2$ of about 85 \%. Varying the gluon field $\chi_0$ by 2\%
would even result in a 
change of $\chi^2$ by many orders of magnitude.
The like is true for $g_4$, $k_1$ and $k_3$.
In contrast $f_{\mathrm K}$, entering the symmetry breaking part of the 
Lagrangian, Eq.\ref{mflagrdichte},
does not change considerably.
In general we see that the parameters change appreciably due to the fit.\\

\subsection{Known Nuclei}
\subsubsection{The Change of the Fit-Observables}
Table \ref{observablen} shows the absolute and relative change of the
parameters entering the fit, due to the adaption of the parameters.
The fit improves the results by more than a factor of 20.
For all nuclei the binding energy has changed most compared to
the original nuclear matter fit $C_1$.
The diffraction radius changes little, for lead there is even a slightly 
worse result than in the fit to infinite nuclear matter.
\subsubsection{The Formfactor}
\begin{figure}[h]
\vspace*{-0.3cm}
\hspace{0cm} 
\psfig{file=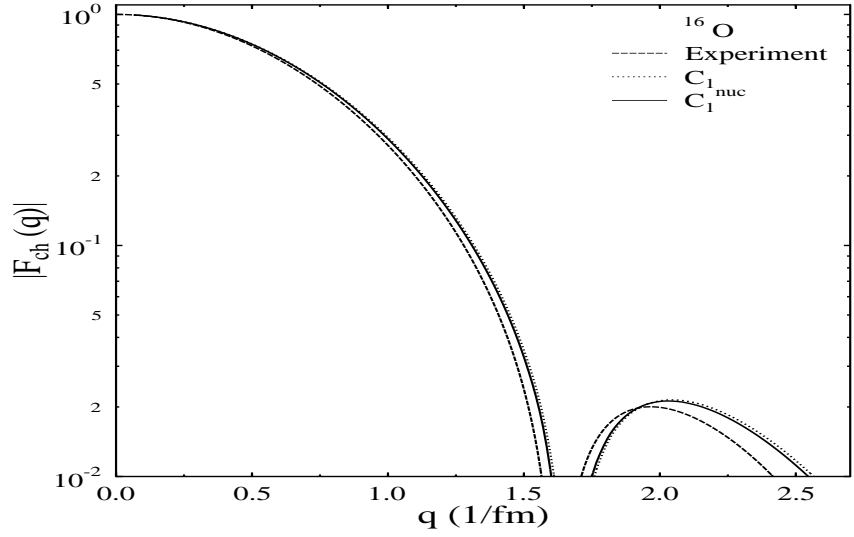, height=9cm ,width=14cm}
\caption{\label{formo}Charge Form Factor of $^{16}$O.}
\end{figure}

\begin{figure}[h]
\vspace*{-0.3cm} 
\hspace{0cm}
\psfig{file=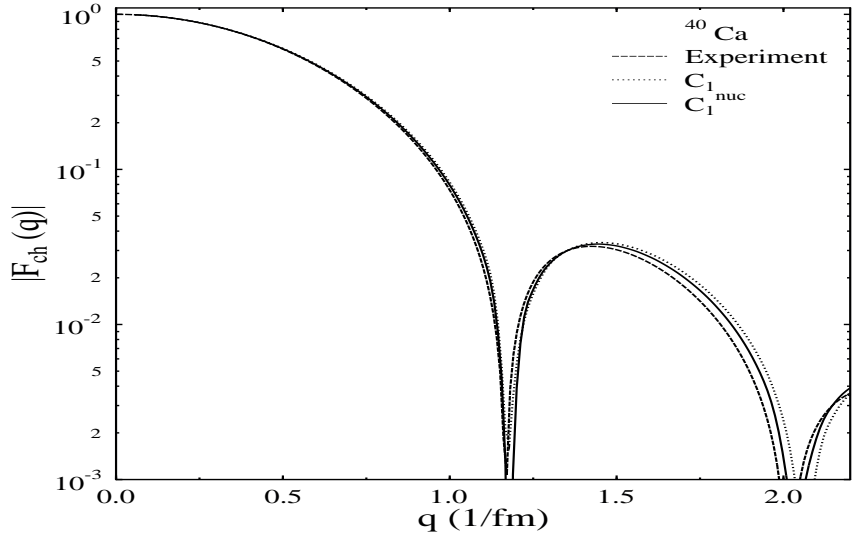, height=9cm ,width=14cm}
\caption{\label{formca}Charge Form Factor of $^{40}$Ca.}
\end{figure}

\begin{figure}[h]
\vspace*{-0.3cm}
\hspace{0cm}
\psfig{file=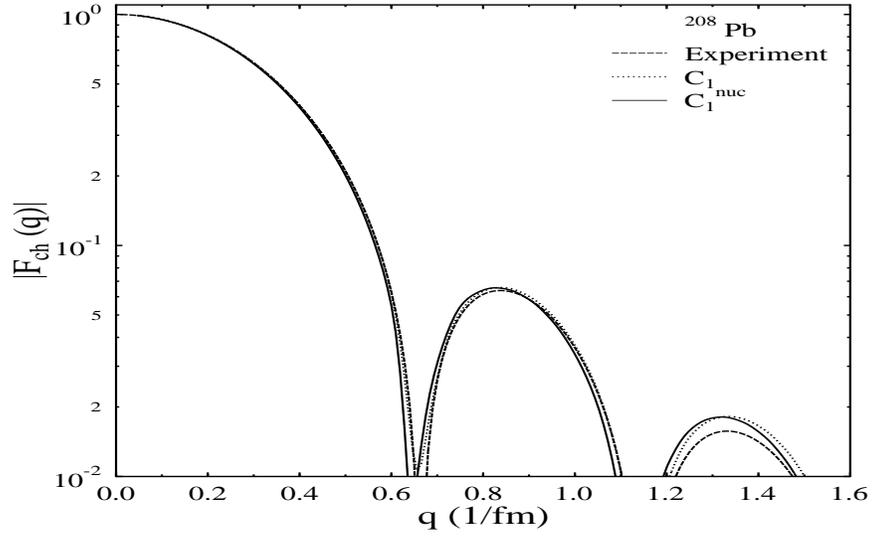, height=9cm ,width=14cm}
\caption{\label{formpb}Charge Form Factor of $^{208}$Pb.}
\end{figure}

We now want to employ our model to finite nuclei with known properties.
Their observables enter into the adjustment of the parameters.
Figures \ref{formo}, \ref{formca} and \ref{formpb} show the charge
form factors of oxygen, calcium and lead.
The experimental values have been taken from \cite{vri87}.
Here the position of the first null and the first maximum are of
particular interest, because they determine the charge diffraction radius and
the surface thickness.
We find that both parameter sets lead to very similar results,
but all calculations show moderate deviations from the experiment \cite{vri87}.
The first null in the formfactor of oxygen lies at about 2 \% 
higher momentum as in the experiment. 
This leads to a slightly smaller nucleus.
Moreover, the surface of the nucleus is too thin, as visible in the
deviation of about 7 \% at the first maximum.
Calcium agrees significantly better with the experiment.\\
For lead the first null is at slightly ($\sim$ 1\%) too small momenta, 
which results in a nucleus which is somewhat too big.
\subsubsection{The Single Particle Energies for Neutrons}
\begin{figure}[h]
\vspace*{-0.3cm}
\hspace{0cm}
\psfig{file=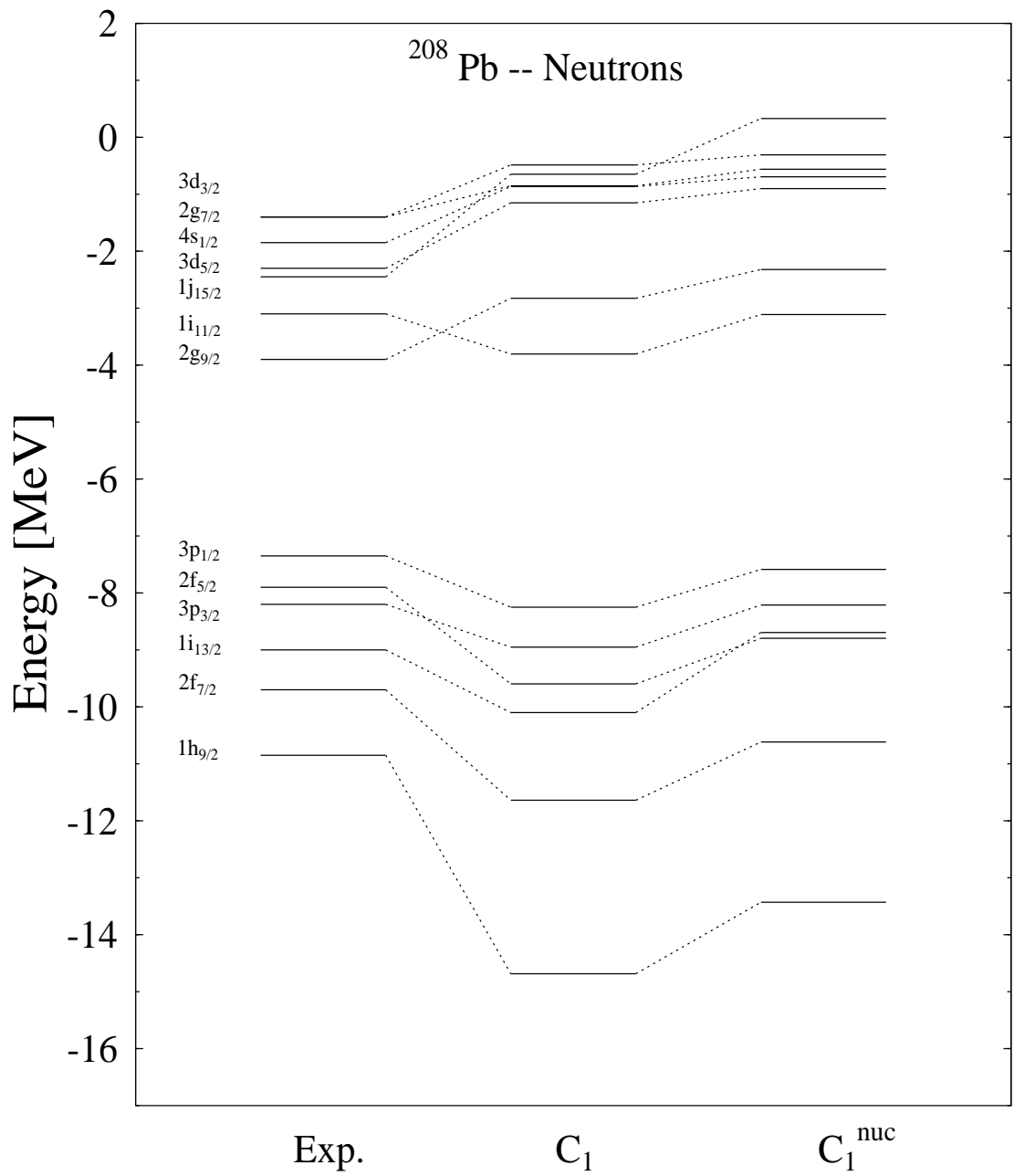, height=20cm ,width=16cm}
\caption{\label{speneu} Single Particle Energies of Neutrons in $^{208}$Pb.}
\end{figure}

Figure \ref{speneu} shows the single-particle energies for neutrons in $^{208}$Pb
(The experimental values are taken from \cite{campi}).
We find a good reproduction of the spin-orbit splitting while the absolute
positions of the energy-levels deviate significantly from the
experimental values.\\
Adapting the parameters to finite nuclei ($C_1^{\mathrm nuc}$) results in 
a shift of the energy-levels, which is obvious because the total
binding energy enters into the fit and experiences a noticeable change
(See Table \ref{observablen}).
It is remarkable that the relative distances between the levels are scarcely
changed. The single particle energies are mostly shifted together.
Furthermore it was not possible to find a set of parameters with a different
distribution of the energy levels, even by changing single parameters by hand.
\subsubsection{Shell Closures and Magic Nuclei}
\begin{figure}[h]
\vspace*{0cm}
\hspace{-0cm}
\psfig{file=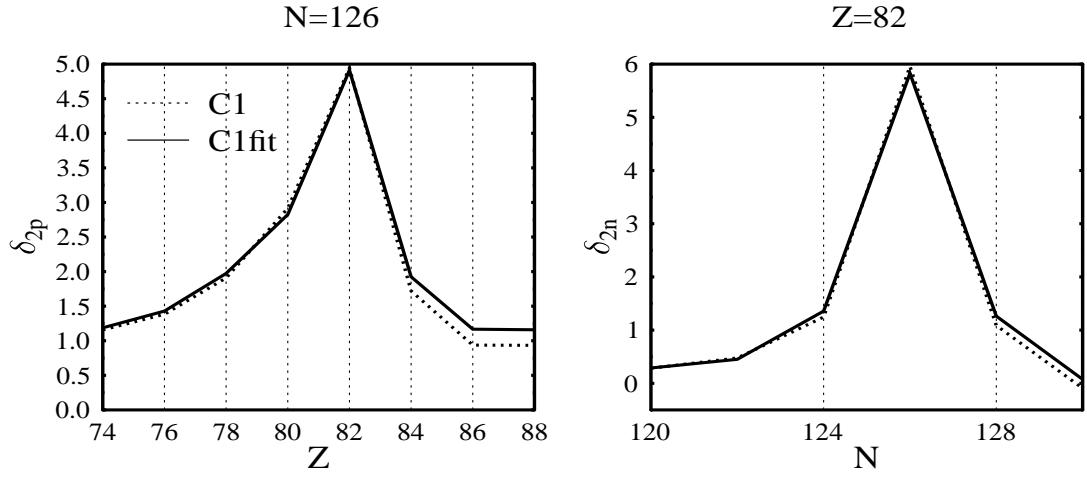, height=17cm ,width=16cm,angle=270}
\caption{\label{magicpb} Two Nucleon Gap-Energy (in MeV) in $^{208}$Pb for Protons and Neutrons.}
\end{figure}

The model reproduces all known magic numbers correctly.
Exemplarily Fig.\ \ref{magicpb} shows the two-nucleon gap energy
\begin{eqnarray}
    \delta_{\mathrm 2n} (Z,N) &=& S_{\mathrm 2n} (Z,N) -S_{\mathrm 2n}(Z,N+2) \\
    \delta_{\mathrm 2p} (Z,N) &=& S_{\mathrm 2p} (Z,N)  - S_{\mathrm 2p}(Z+2,N)  
\end{eqnarray}
with the two-nucleon separation energy
\begin{eqnarray}
    S_{\mathrm 2n} (Z,N) &=& E_{\mathrm B}(Z,N-2) - E_{\mathrm B}(Z,N) \\
    S_{\mathrm 2p} (Z,N) &=& E_{\mathrm B}(Z-2,N) - E_{\mathrm B}(Z,N) \q . 
\end{eqnarray}
The binding energy is given by
\begin{equation}
E_{\mathrm B}=E_{\mathrm total}-A m_{\mathrm N} \q .
\end{equation}
The peaks indicate the magic numbers, in this case $Z=82$ for protons
and $N=126$ for neutrons.\\
Fitting the model parameters to finite nuclei instead
of infinite nuclear matter does not change the
magic numbers. This also holds for all other known magic nuclei.
It should be noticed that of course the binding enegies per nucleon
do change with the fit. 
\subsection{Superheavy Nuclei}
The investigation of nuclei with more than 100 protons 
has recently received renewed interest. After the elements 110 and 112 
\cite{hof95,hof96}, produced at GSI
in Darmstadt, remained the heaviest isotopes for several years, 
it was possible to produce nuclei with 114 and 116 protons at Dubna
in 1999 \cite{oga99}, a nucleus with 118 protons 
might have been observed in a Berkeley experiment \cite{nin99}.
The quest for superheavy nuclei is motivated
by the prediction of shell closures
beyond $Z$=110. In the past such calculations have already been done in 
relativistic mean field models, for example \cite{rut97,ben99}
predicting $Z$=114 and $Z$=120 and \cite{cwiok} predicting $Z$=126.
In the following we want to present the results of the chiral
model developed here.
\subsubsection{Shell Closures}
\begin{figure}[h]
\vspace*{-0.3cm}
\hspace{0cm}
\psfig{file=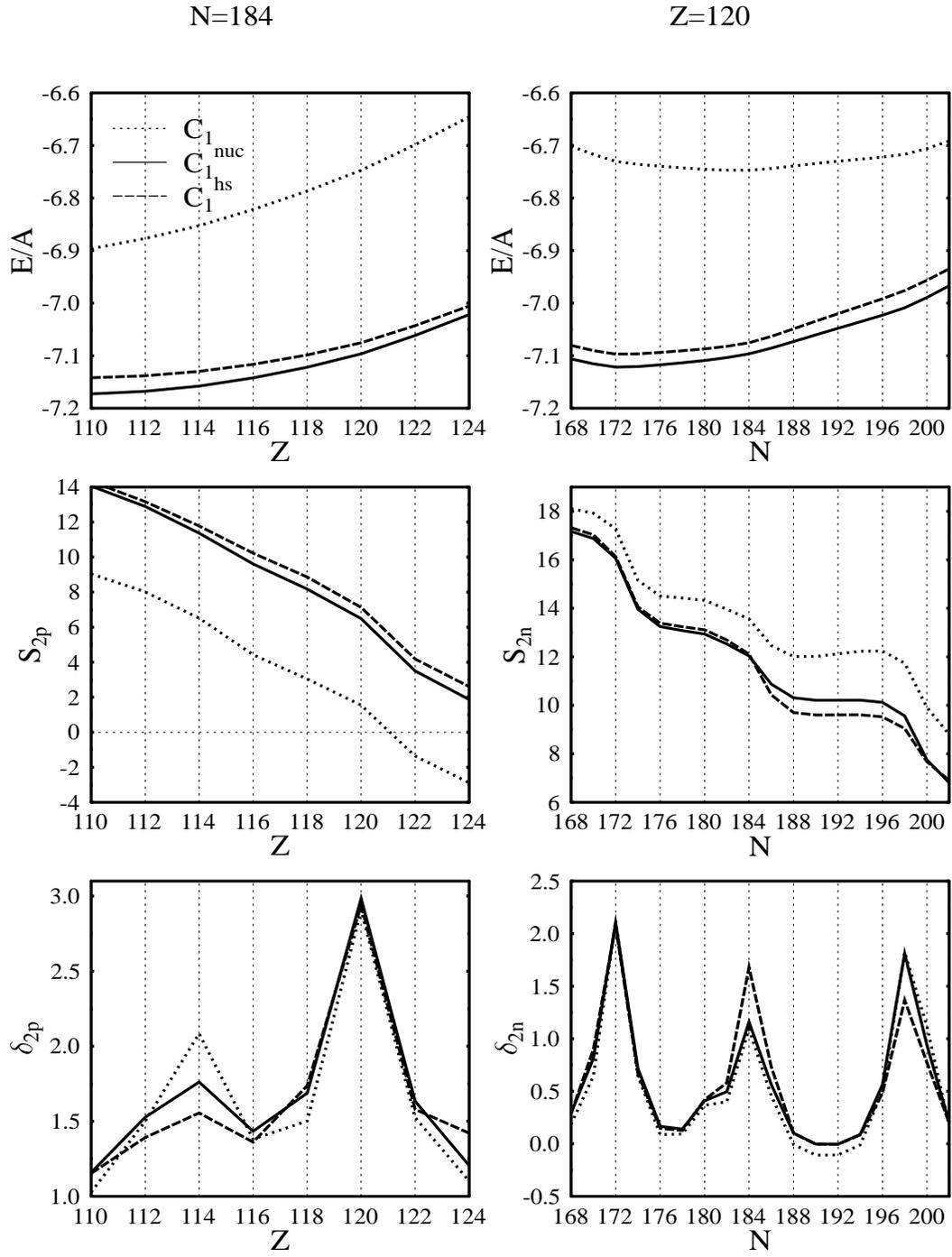, height=20cm ,width=16cm}
\caption{\label{n184z120} Binding energy, two nucleon seperation energy and
two nucleon gap energy for nuclei with 184 neutrons and 120 protons (in MeV).}
\end{figure}

The chiral model predicts clearly $Z$=120 as the next shell closure,
while another one at $Z$=114 is visible but weak.
Fig.\ \ref{n184z120} shows the binding energy, the 
two nucleon separation energy and the two-nucleon gap for nuclei with
184 neutrons and varying proton numbers (left side) and
for nuclei with 120 protons and varying neutron numbers (right side).
The parameter sets used here are explained above.
The major change happens with the use of parameters fitted to finite nuclei
instead of infinite nuclear matter.
The most important point to mention here is the suppression of
the peak in the two-proton gap at the proton number of $Z$=114.
We find that it is clearly smaller in the calculation with $C_1^{\mathrm nuc}$
compared to $C_1$ and it is even more suppressed when using the set $C_1^{\mathrm hs}$.

In conclusion we find that $Z$=114 is at best a subshell closure 
in this chiral model, 
while $Z$=120 is probably the next shell closure which
shows a distinctly pronounced peak.
Furthermore it should be mentioned that this nucleus is still above the
drip line which shifts to higher $Z$ in the fit to nuclei compared to 
the nuclear matter fit.\\
For neutrons we find the next magic numbers at $N$=172, $N$=184 and
$N$=198.\\
\subsubsection{Charge Distribution and $\alpha-Decay$}
\begin{figure}[h]
\vspace*{-0.3cm}
\hspace{0cm} 
\psfig{file=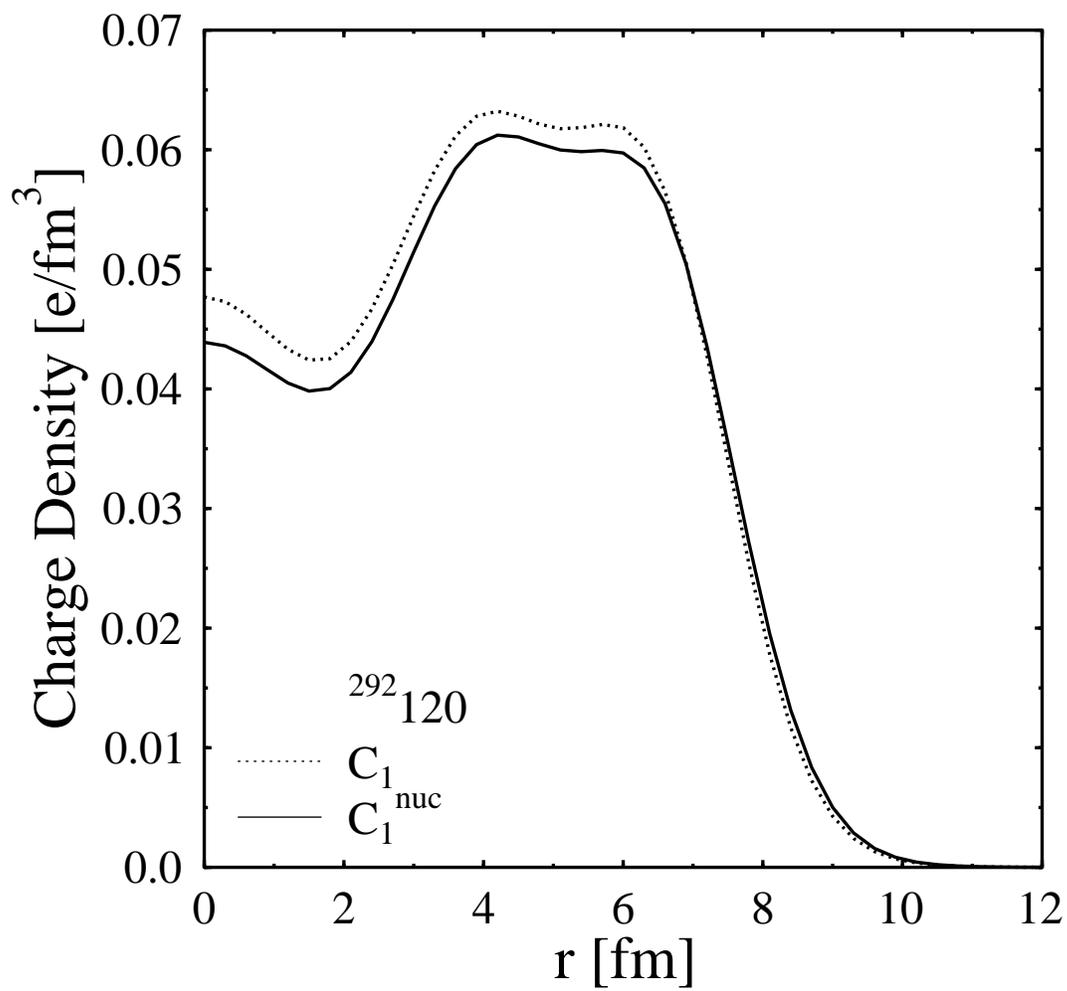, height=16cm ,width=16cm}
\caption{\label{ladungsvert} Charge distribution in the nucleus $^{292}_{172}$120.}
\end{figure}

\begin{figure}[h]
\vspace*{-0.3cm}
\hspace{0cm}
\psfig{file=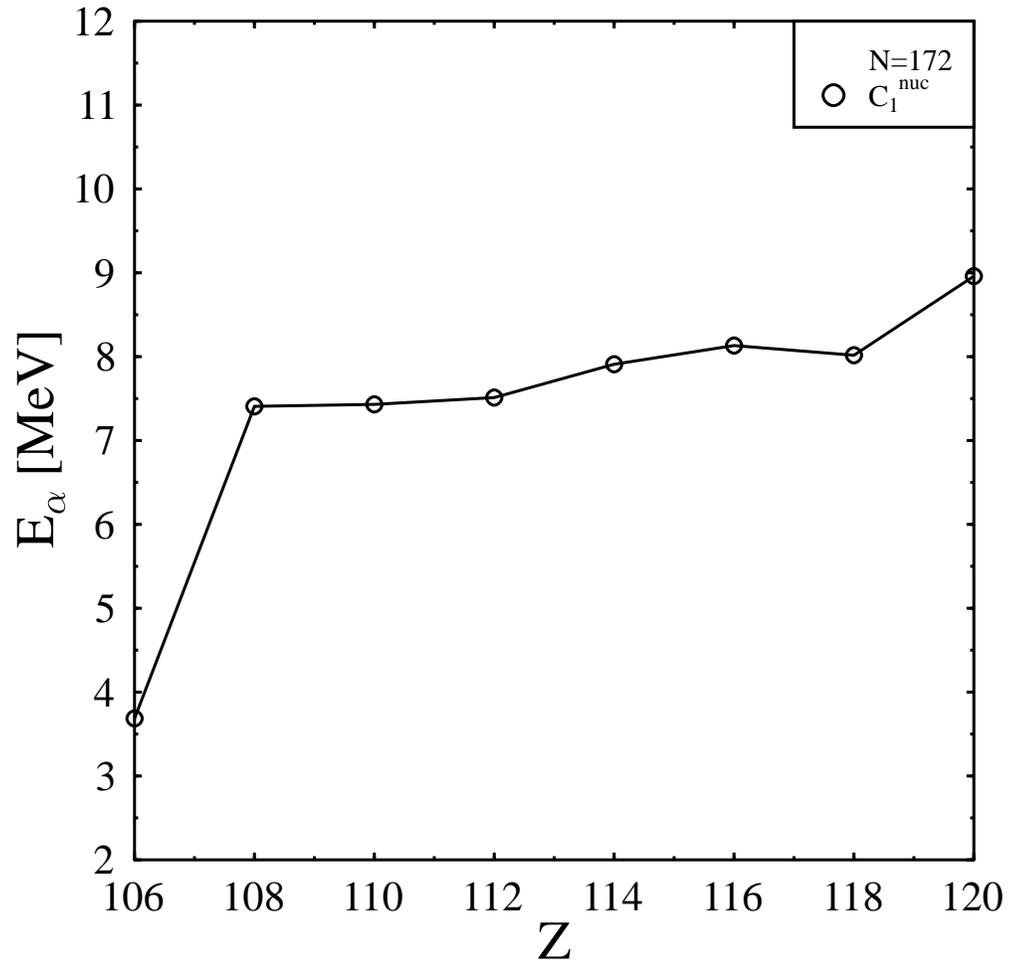, height=16cm ,width=16cm}
\caption{\label{alpha} Energies of the $\alpha$-particles emmited
in the decay of the nucleus $^{292}$120.}
\end{figure}

A specialty of superheavy nuclei seems to be their charge distribution
which is shown in Fig.\ \ref{ladungsvert} for the model in question.
One finds a noticeable decrease of the charge density in the center of
the nucleus. This observation has also been made in calculations
in the Walecka model \cite{ben99}.
Could this behaviour be the result of a structure of the nucleus, which
reminds of the bucky balls in organic chemistry ?
The nucleus in question has 120 protons what corresponds to 
60 $\alpha$-particles. These could form a bucky ball structure
hold together by valence neutrons.
Lighter objects of this type have been investigated earlier by
W.\ von Oertzen \cite{oertzen}.\\
The characteristic observables for the identification of
superheavy nuclei are the energies of the $\alpha$-particles emitted during
the decay. The resulting energies in the chiral model are
shown in Fig.\ \ref{alpha}.
One finds that these energies are almost constant in the region
from $Z$=118 to $Z$=108. This is in {\it qualitative} agreement with 
the possible
findings of the Berkeley group which found the nucleus $^{293}118_{175}$.
At this point one should keep in mind of course that our
calculations are all done in spherical approximation.
\subsection{Hypernuclei}
\begin{figure}[h]
\vspace*{-0.3cm}
\hspace{-3cm}
\epsfig{file=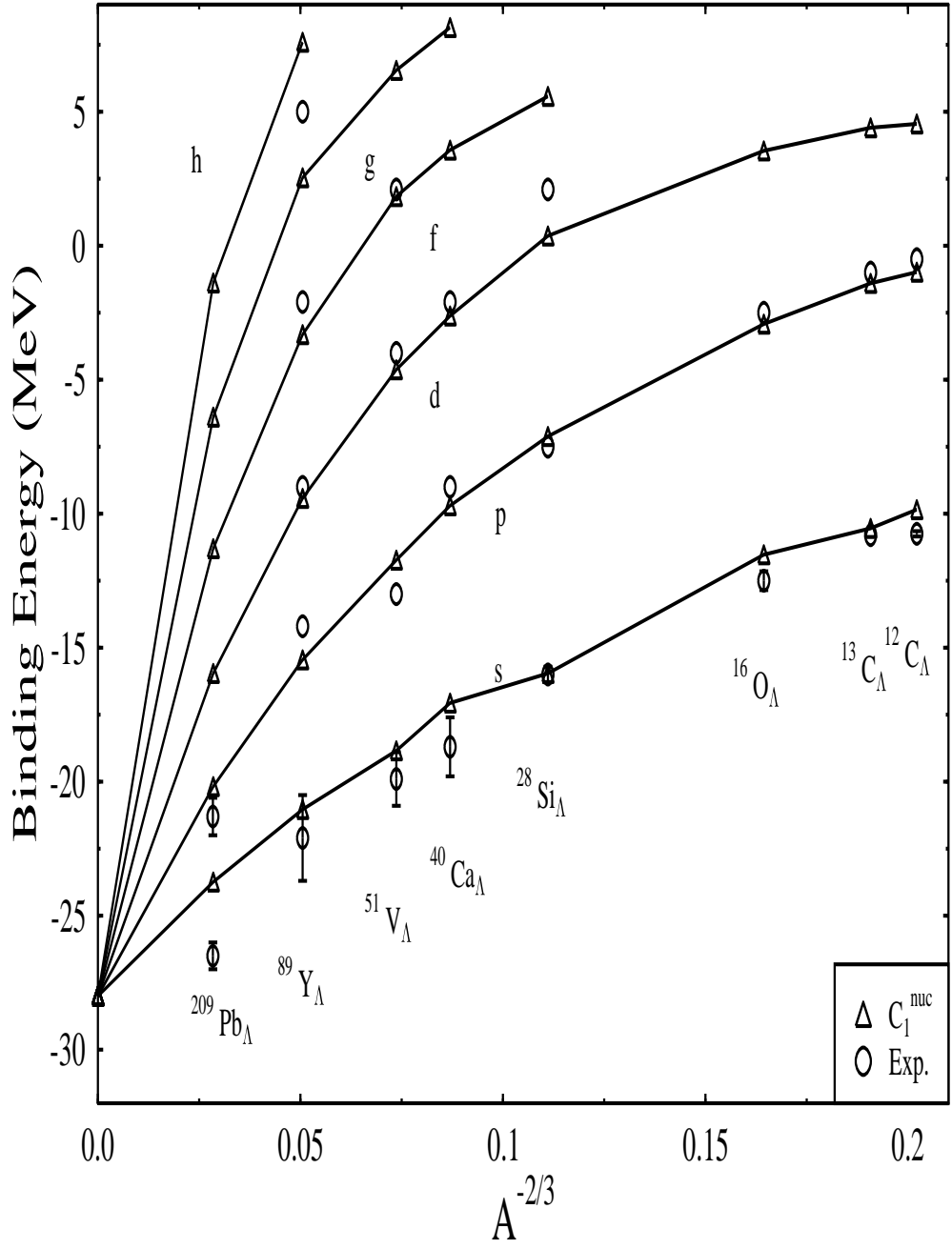, height=22cm ,width=17cm}
\caption{\label{lambniv} Single-particle energies of $\Lambda$-particles
in different nuclei.}
\end{figure}

\begin{figure}[h]
\vspace*{-0.3cm}
\hspace{-3cm}
\epsfig{file=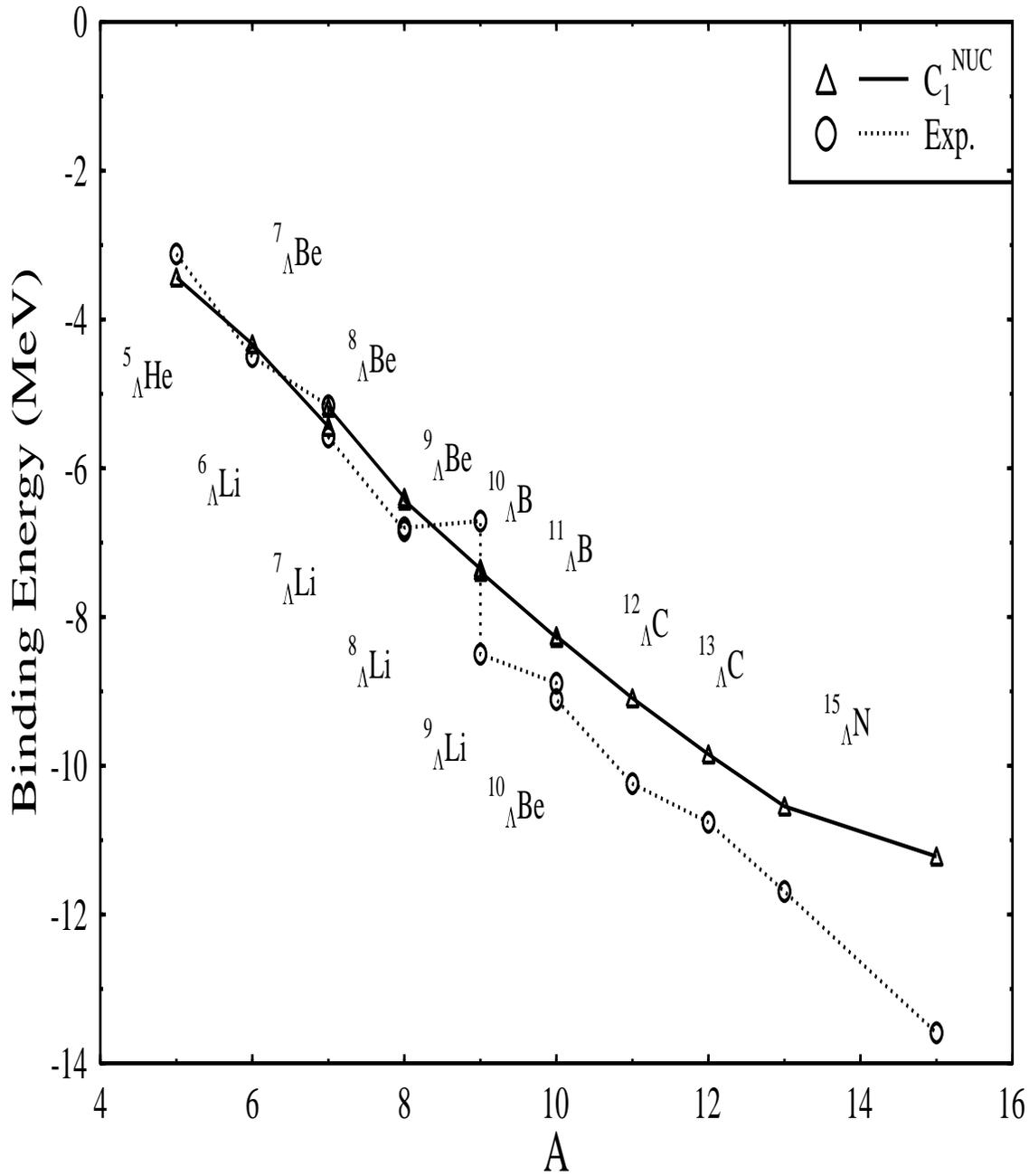, height=22cm ,width=17cm}
\caption{\label{lambdaleicht} Binding Energy of $\Lambda$-baryons in leight
nuclei.}
\end{figure}

\begin{figure}[h]
\vspace*{-0.3cm}
\hspace{-3cm}
\epsfig{file=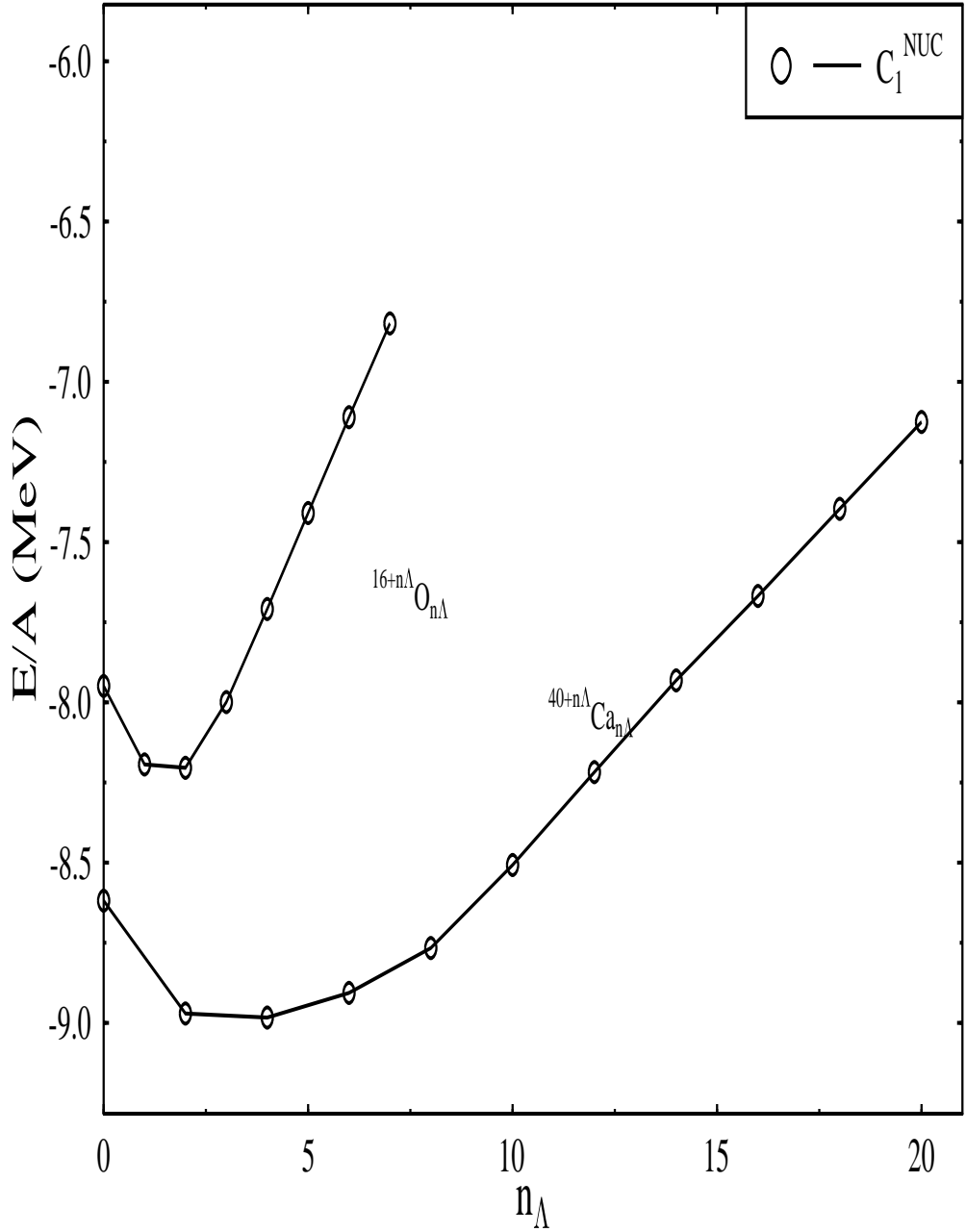, height=22cm ,width=17cm}
\caption{\label{multilamb} Binding Energy of nuclei with different 
numbers of added hyperons.}
\end{figure}

We now want to apply the chiral model to nuclei including hyperons.
Hypernuclei have been first observed by Danysz and Paiewski in 
1953 \cite{dan53}. Hypernuclei have a lifetime of about $10^{-10}$s
which makes it particularly difficult to investigate heavy ones.
In the seventies, hypernuclei with up to $A$=15 have been produced
\cite{can74}
and some years later even heavier hypernuclei could be observed at CERN
\cite{bru76,bru78,ber79,ber80,ber81} and AGS \cite{bon74,chr79,may81,chr88}
where also exited states have been investigated.\\
A theoretical description of hypernuclei was tried among others
in Skyrme-Hartree-Fock-models (e.g. \cite{mil88,ine96}) and relativistic 
mean-field models \cite{mar89,ruf88}.\\
In the following calculations of hypernuclei in the chiral $SU(3)$-Model
shall be investigated.\\
Fig.\ \ref{lambniv} shows the single particle energies of $\Lambda$-particles
in different nuclei (Experiment:\cite{aji95,has96}). 
Here the binding energy of $\Lambda$-particles 
is plotted against $A^{-2/3}$, i.e. the inverse of the
surface of the nucleus. The parameters were adapted to finite nuclei 
($C_1^{\mathrm nuc}$). All single particle levels meet for $A\longrightarrow \infty$ 
at the value of the
$\Lambda$-potential (-28 MeV) in nuclear matter as it should be.
However one should keep in mind that it is possible to modify the
depth of the hyperon potential in nuclear matter
by the explicit symmetry-breaking in
strangeness direction without effecting the properties of normal
nuclei \cite{paper3}.
One achievment of the model is the correct description of the $\Lambda$-levels
in finite nuclei, that is the reproduction of the measured 
single-particle-energies, which is particulary successful for the deeper bound
states. An exception is the 1s-niveau in $^{208}$Pb which is 
deeper bound than in our calcuation.
However this experimental result also contradicts a nuclear matter potential
of -28 MeV, because the $\Lambda$-level itself lies at -26.5 MeV
and the determiantion of the nuclear matter potential is done by
extrapolating the $\Lambda$-levels in finite nuclei.
The situation is similar to calculations in 
relativistic mean-field models \cite{sch92,ruf90}, where one 
finds even less deeply bound states.\\
Fig.\ \ref{lambdaleicht} shows the binding energy of $\Lambda$-baryons in 
light nuclei (It was calculated the 1-s-level in a nucleus without $\Lambda$).
The experimental data have been taken from \cite{juric}.
In the region of light nuclei the calculations agree very well with
the experiment, for heavier nuclei the divergence gets larger.
This is especially astonishing since one would expect larger
errors for light nuclei in a mean field calculation.
The substantial point is here that the calculations show a
different trend than the experiments.
Furthermore the energy differences between nuclei of equal masses 
are not reproduced in the calculations.
Anyhow, the result is satisfying since the only hypernucleus observable
entering the model, is the nuclear matter potential of the $\Lambda$.\\
Since the model provides reasonable results for nuclei with only 
one $\Lambda$, it is quite interesting to investigate nuclei with 
more than one $\Lambda$.
However this field is experimentally largely unexplored. The only
nuclei ever observed with more than one $\Lambda$ are 
$^6_{\Lambda\Lambda}$He, $^{10}_{\Lambda\Lambda}$Be and 
$^{13}_{\Lambda\Lambda}$Be \cite{dan63,pro66,aok91}.\\
Fig.\ \ref{multilamb} shows the calculated binding energies for 
$^{16}$O and $^{40}$Ca with a variing number of added 
$\Lambda$-particles.
One finds minima at non-zero numbers of added $\Lambda$ for
oxigen and calcium. This is not surprising because with the
$\Lambda$-particles a new degree of freedom has been opened.
The $\Lambda$-potential is below
the highest occupied nucleonic state.\\
All shown nuclei are stable, because the 
target states of the decay are occupied with nucleons (Pauli-Blocking).\\
The basis for the good description of $\Lambda$-hypernuclei is 
given by the
correct value of the $\Lambda$-potential in inifinite nuclear matter.
We will see in the next section that this is
not fulfilled for the other hyperons which leads to problems with the
investigation of e.g. $\Xi$-matter.
\subsection{MEMOS}
While experimental results for nuclei with one hyperon exist for 
some time, there is almost nothing known about objects with 
a higher amount of strangeness.  
We now want to investigate, if nuclear objects containing 
$\Xi$-hyperons (metastable, exotic multistrange objects (MEMOs))
are stable against strong decay in the investigated model.
Experimentally no strange objects, additional to those mentioned above,
are known.\\
Theoretically MEMOs were investigated copious in Walecka-type
mean-field-models \cite{sch92,sch93,sch94a,sch94b}.
There in fact the authors find possibilities of combining hyperons,
which lead to metastable objects.\\
In the current chiral model substantial difficulties occure due to the
fact that the hyperon potentials are not described correctly.\\
For the correct description of the $\Lambda$-potential in nuclear matter
an explicit symmetry breaking term has been included, which
does not violate the PCAC-relation \cite{pap98}:
\begin{equation}
\label{hyesb}
  {\cal L}_{{{\mathrm hyp}}} = m_3 {\mathrm Sp} \li(\overline{B} B + 
\overline{B} [B, S]\re) {\mathrm Sp}(X-X_0) 
\end{equation}
with $S_{b}^a = -\frac{1}{3}[\sqrt{3}(\lambda_8)_{b}^a-\delta_{b}^a]$.
The parameter $m_3$ is used to fix the $\Lambda$-potential at -28 MeV.
The other hyperon potentials are determined by this.
This leads to a repulsive $\Xi$-potential of 30.3 MeV.
The potential of $\Xi$-particles in $\Xi$-matter is about 126 MeV at 
nuclear density and thus strongly repulsive as shown in Table \ref{mitsb}.
One also finds that the $\Lambda$-potential in $\Lambda$-matter is
repulsive.\\
In summary
there is no combination of $\Lambda$- or $\Xi$-particles which
would lead to bound states. The only stable combinations are given by
$\Lambda$-particles in nuclear matter as shown in the previous section.
Furthermore the model predicts that $\Xi$-hypernuclei cannot exist
because the nucleon-$\Xi$-potential is repulsive, too.\\
If one neglects the explicite symmetry breaking in the strange sector
completely the models yields attractive potentials between hyperons 
(Table \ref{ohnesb}), but on the other hand the model is not able to reproduce 
a single hyperon-observable correctly. For example the potential of 
$\Lambda$-particles in nuclear matter is then $U_\Lambda$=-100 MeV which is
far too deep. The same is true for the $\Xi$-potential with $U_\Xi$=-115 MeV.\\
Another possibility to make the existance of MEMOs possible 
is to include a further parameter to the symmetry breaking. 
In this case it is possible to make the hyperon potentials
a bit less repulsive, but nevertheless they stay repulsive, so MEMOs
can not exist (Table \ref{zuspara}).
Admittedly nuclei are possible in this case which include nucleons together
with $\Xi$-hyperons, because their potentials in nuclear matter are 
attractive.
For example a $^4$He nucleus with maximum two $\Xi$-particles is 
imaginable, in $^{16}$O even four bound $\Xi$-states could be possible if
one takes into account the change of the potential due to the
added hyperons.
It should nevertheless be mentioned that all these nuclei would not
be stable against strong decay, because the nucleon states, in those the
hyperons would decay are not occupied.
 
\subsection{Conclusion}
A chiral $SU(3)$-model has been applied to different forms of
finite nuclear matter.
The model is based on chiral symmetry in nonlinear realisation.
All calculations have been performed in mean-field-approximation
and spherical symmetry. The parameters have been fitted to properties of
finite nuclei.\\
The model shows a clear improvement of the results compared to 
calculations with parameters fitted to infinite nuclear matter.
All charge distributions of spherical nuclei can be reproduced 
satisfactorily. Furthermore all known shell closures are described
correctly.
The qualitative results are very robust against changes of the parameters.\\
In the regime of superheavy nuclei the model predicts clearly $Z$=120 as the
next magic number. The closure at $Z$=114, which is often shown in 
non-relativistic models, is not validated by our model.
For the neutrons the model yields $N$=172, $N$=184 and $N$=198 as magic 
numbers.
These results are widely independent of the parameter set.
For $Z$=120 the charge distribution shows a strong depletion
in the center of the nucleus.\\
Hypernuclei have also been investigated.
The potential of the $\Lambda$-particles could be fixed by the explicit
symmetry breaking term. One then observes a very good reproduction of
the experimental data.
In particular some nuclei with one or more $\Lambda$-particles are
resistent against strong decay.\\
Major problems appear with objects containing $\Xi$-hyperons 
or those that contain no normal nucleons at all (MEMOs).
Because all hyperon-hyperon-potentials are repulsive,
there is not a single bound hypernucleus built up from hyperons alone.
Furthermore the potential of $\Xi$-particles in nuclear matter is 
positive.\\
\subsection{Outlook}
Up to now all calculations have been performed in spherical symmetry.
Specially for superheavy nuclei it is most interesting to investigate
the influence of deformation.
A further investigation of the density-depletion in the center of $Z$=120
is desirable, particularly under the aspect of cluster formation.
The question about the existence of MEMOs has to be investigated further.
The preferable solution would be to find a term 
which describes all hyperon potentials
correctly but does not introduce new parameters which have to be
adjusted.
Possibly one will have to change the model on more basic level and
has to introduce an additional scalar condensate, which is not the 
chiral partner of the pion.
Investigations along this line are in progress.
\subsection{Aknowledgements}
This work was supported by GSI, BMBF, DFG and the Josef Buchmann 
Stiftung.


\begin{table} 
\begin{center}
\begin{tabular}{c|c|c|c|c}
   &$C_1$   & $C_1^{\mathrm nuc}$ & $C_1-C_1^{\mathrm nuc}$ & $C_1-C_1^{\mathrm nuc}$ [\%]\\ 
\hline
$g_{{\mathrm N}\omega}$&13.6065&13.5723&0.0342&0.25\\
$\chi_0$     &401.93 &409.77 &-7.84 &-1.95\\
$f_{\mathrm K}$        &122.0  &122.143&-0.143&-0.12\\
$g_4$        &61.47  &74.57&-13.1054&-21.32\\
$k_1$        &1.399  &1.354&0.0456  &3.256\\
$k_3$        &-2.6525&-2.773&0.12&-4.542\\
$g_{{\mathrm N}\rho}$  &4.5355&5.6579&-1.1224&-24.75
\end{tabular}
\end{center}
\caption{\label{parameter} The parameters used with the model and
their change due to the fit to finite nuclei.}
\end{table}
\begin{table}
\begin{center}
\begin{tabular}{c|ccc|ccc|ccc|c}
      & \multicolumn{3}{c}{$^{16}$O} & \multicolumn{3}{c}{$^{40}$Ca} &\multicolumn{3}{c}{$^{208}$Pb}\\ 
      &      E/A (MeV) & $R_{\mathrm{diff}}(fm)$ & $\sigma_{\mathrm O} (fm)$ & E/A & $R_{\mathrm{diff}} $ &$\sigma_{\mathrm O} $ & E/A  & $R_{\mathrm{diff}} $  
 & $\sigma_{\mathrm O} $ & $\chi^2$ \\ \hline
Exp.  
& -7.98    &  2.78     & 0.84
& -8.55  & 3.85 & 0.98  
& -7.86  & 6.81 & 0.90 &  \\
$C_1$ 
& -7.30   &  2.68    &0.79 
&-8.00   &  3.80 & 0.92 
& -7.56 &    6.79   &  0.87& 7749 \\
$C_1^{\mathrm nuc}$ 
& -7.95   &  2.7    &0.81 
&-8.62   &  3.81 & 0.94 
& -7.91 &    6.86  & 0.89 & 361
\end{tabular}
\end{center}
\caption{\label{observablen} The values of the fit observables for paramters adapted to
infinite nuclear matter ($C_1$) as well as finite nuclei ($C_1^{\mathrm nuc}$).} 
\end{table}

\begin{table} 
\begin{center}
\begin{tabular}{c|c|c|c}
&N&$\Lambda$&$\Xi$\\
\hline
N-matter&-71.04&-28.23&+30.27\\
$\Lambda$-matter&-38.13&+20.45&+85.78\\
$\Xi$-matter&+16.17&+73.83&+126.75
\end{tabular}
\end{center}
\caption{\label{mitsb}Baryon potentials (in MeV) in different environments including
explicit symmetry breaking.}
\end{table}

\begin{table} 
\begin{center}
\begin{tabular}{c|c|c|c}
&N&$\Lambda$&$\Xi$\\
\hline
N-matter&-71.04&-100.6&-114.46\\
$\Lambda$-matter&-97.31&-89.2&-73.27\\
$\Xi$-matter&-126.94&-90.57&-56.59
\end{tabular}
\end{center}
\caption{\label{ohnesb} Baryon potentials (in MeV) in different environments
without explicit symmtry breaking.}
\end{table}

\begin{table}
\begin{center}
\begin{tabular}{c|c|c|c}
&N&$\Lambda$&$\Xi$\\
\hline
N-matter&-71.04&-28.23&-42.09\\
$\Lambda$-matter&-38.13&+20.45&+30.68\\
$\Xi$-matter&-60.16&+21.06&+49.43
\end{tabular}
\end{center}
\caption{\label{zuspara} Baryon potentials (in MeV) with a symmetry-breaking,   
including an additional parameter.}
\end{table}


\end{document}